\documentclass[english,aps,pra,twocolumn]{revtex4}
\usepackage[T1]{fontenc}
\usepackage[latin9]{inputenc}
\usepackage{CJK}
\usepackage{mathrsfs}
\usepackage{amsmath}
\usepackage{amssymb}
\usepackage{graphicx}
\usepackage{epstopdf}
\usepackage{subfigure}
\usepackage{color}
\usepackage{ulem}

\makeatletter

\@ifundefined{textcolor}{}
{%
 \definecolor{BLACK}{gray}{0}
 \definecolor{WHITE}{gray}{1}
 \definecolor{RED}{rgb}{1,0,0}
 \definecolor{GREEN}{rgb}{0,1,0}
 \definecolor{BLUE}{rgb}{0,0,1}
 \definecolor{CYAN}{cmyk}{1,0,0,0}
 \definecolor{MAGENTA}{cmyk}{0,1,0,0}
 \definecolor{YELLOW}{cmyk}{0,0,1,0}
}

\usepackage{babel}

\makeatother

\usepackage{babel}
\begin{document}
\title{The effect of static disorder on the center line slope in 2D electronic spectroscopy}
\author{Zong-Hao Sun}
\affiliation{Department of Physics, Applied Optics Beijing Area Major Laboratory, Beijing Normal University, Beijing 100875, China}

\author{Yi-Xuan Yao}
\affiliation{Department of Physics, Applied Optics Beijing Area Major Laboratory, Beijing Normal University, Beijing 100875, China}


\author{Qing Ai}
\email{aiqing@bnu.edu.cn}
\affiliation{Department of Physics, Applied Optics Beijing Area Major Laboratory, Beijing Normal University, Beijing 100875, China}

\author{Yuan-Chung Cheng}
\email{yuanchung@ntu.edu.tw}
\affiliation{Department of Chemistry and Center for Quantum Science and Engineering, National Taiwan University, Taipei City 106, Taiwan}

\begin{abstract}
Two-dimensional electronic spectroscopy (2DES) is a powerful tool for investigating the dynamics of complex systems. However, analyzing the resulting spectra can be challenging, and thus may require the use of theoretical modeling techniques. The center line slope (CLS) method is one of such approaches, which aims to extract the time correlation function (TCF) from 2DES with minimal error. Since static disorder is widely observed in complex systems, it may be interesting to ask whether the CLS approach still work in the presence of the static disorder. In this paper, the effect of the static disorder on the TCF obtained through the CLS method is investigated. It is found that the steady-state value of the CLS increases monotonically with respect to the static disorder, which suggests that the amplitude of the static disorder can be determined using the CLS in the long-time limit. Additionally, as the static disorder rises, the decay rate of the CLS first decreases to a certain value and remains at this value until the static disorder is sufficiently large. Afterward, the CLS begins to fluctuate significantly and thus results in obtaining the decay rate through the CLS method unreliable. Based on these discoveries, we propose a method to fix the error and obtain the TCF. Our findings may pave the way for obtaining reliable system-bath information by analyzing 2DES in the practical situations.

\end{abstract}
\maketitle

\section{introduction}

2DES is a powerful spectral technology developed in recent years for analyzing the dynamics of a variety of chemical and biological systems \cite{collini2009sci,song2014NC,de2016NC,harel2012Pro,fidler2013JPCL,ostroumov2013sci,fidler2014NC}. It has high resolution in both time and frequency domains, and has been successfully applied to probing fast dynamics in condensed-matter systems with exceptional detail.
The 2DES is a branch of the two-dimensional spectroscopy in the visible domain, which is widely used in the study of photoactive systems including photosynthetic complexes \cite{collini2010NA,calhoun2009jpcb,panitchayangkoon2010pro}, photovoltaics \cite{monahan2017jpcl,richter2017NC}, nanocrystalline \cite{turner2012nano,cassette2015na,stoll2017jpcl}, quantum dots and wells \cite{caram2014jpcl,nardin2014prl,park2017nano}, and photosynthetic pathways \cite{romero2014np,duan2017scire}. The nonlinear broadening, energy transfer, electron-coupling effects and quantum coherence effects can be intuitively demonstrated in the 2DES. Several closely-related technologies have also been developed, such as 2D fluorescence spectroscopy \cite{perdomo2012jpcb,maly2018jpcl,liang2021jcp}, and 2D terahertz \cite{kuehn2011prl,kuehn2011jpcb}.

In the 2DES experiments, three ultrafast pulses covering the frequency domain of interest successively pass the sample. By controlling the delay times of the pulses, a photon echo signal is emitted in the direction of phase matching after the interaction between three pulses and the sample. The photon-echo signal is combined with another local-oscillator signal for heterodyne detection, which provides the amplified signal for the quantum dynamics.

The first coherence period $\tau$ is the duration between the first two pulses. The population period $T_w$ is the duration between the second and third pulses. The second coherence period $t$ is the duration between the 3rd pulse and the signal. They can be effectively adjusted by tuning the delay times of the pulses.
The electrons are labeled by frequency during the first coherence period. Due to the microscopic events that occur during the population period, the frequency-labeled electrons may develop to various frequencies, which is called spectral diffusion. The final frequencies of the electrons with frequency labels are read out during the second coherence period. By taking the initial frequency which labels the electrons as one axis and the final frequency as the other axis, a 2D spectrum can be obtained. The detailed information and quantum dynamics of the system can be determined by analyzing the position, amplitude and shape of the peaks in the 2D spectrum.

2DES contains immense information about the system. It is crucial how to interpret the spectrum to obtain the required information. In order to explore the dynamic evolution of the system, we focus on extract the TCF accurately and efficiently from the two-dimensional spectrum. The TCF provides a key connection between 2DES photon-echo experiments and microscopic dynamics. Hence, many methods have been developed, such as the CLS \cite{Kwak2007JCP,Kwak2008JCP}, ellipticity and eccentricity \cite{finkelstein2007proc,fang2008proc}.

Among methods for extracting information describing the system-bath interaction, the CLS theory yields reliable TCF and has been successfully applied to describe two-dimensional infrared
vibrational echo spectroscopy and structural
dynamics under thermal equilibrium.
The CLS is the slope (the inverse of the slope) of the center line that connects the peaks of a series of cuts through the 2D spectrum parallel to the $\omega_{t}$ ($\omega_{\tau}$) axis. As spectral diffusion progresses, the CLS decays from a maximum of 1 to 0. The CLS is used as the TCF to study the ultrafast dynamics of the system.

The CLS theory is developed based on the theory of optical response function \cite{Mukamel1999}. In the derivation, the TCF is assumed as a real function, and many approximation methods have been applied, such as the short-time approximation. Furthermore, all pulses are regarded as delta pulses, which may significantly deviate from the practical situation.

In natural photosynthetic complexes, the TCF is complex and can only be considered as a real function at the high-temperature limit \cite{Mukamel1999}. As a result, it might be crucial to test the reliability of the CLS approach when the TCF is a complex function. On the other hand, due to the heterogeneity, the inhomogeneous broadening is introduced due to the static disorder and it may effectively prolong the coherence signals in the 2DES \cite{Dong2016JPCL,Dong2014jpcb}. Therefore, it might be important to investigate the effect of
the static disorder on the CLS. Interestingly, it is found that within a certain range of the static disorder, the performance of the CLS approach has been improved in the extracting the TCF.

In the next section, we give a brief introduction to the optical response function for 2DES and the CLS method. In Sec.~\ref{Sec:SimulatedSpectrum}, we examine the performance of the CLS method at low temperatures. Using complex TCF, we generate 2D spectra at 77~K and 298~K. In Sec.~\ref{Sec:StaticDisorder}, we calculate 2DES with static disorder and investigate its impact on the CLS method. In Sec.~\ref{Sec:Dicussion}, we summarize our main findings. In Appendix~\ref{App:Distribution}, we provide a brief derivation to the probability of the energy gap in the presence of the static disorder. In Appendix~\ref{App:TCF}, we derive the relationship between the real and imaginary parts of the TCF.

\section{methods}
\label{Sec:methods}

The 2DES is obtained from the double Fourier transform of the third-order macroscopic polarization signal generated by three laser pulses acting on the system. The 2DES is calculated by the response function approach. Afterwards, the center line of the 2DES is obtained and the time correlation function is reproduced by the slope of the center line. Hereafter, we shall summarize the two approaches respectively.

\subsection{ Response function and center line slope}

Response function is the most commonly-used method for calculating
the 2DES, which was developed by S. Mukamel and his collaborators
\cite{Mukamel1999}. The two-dimensional spectroscopy can be determined
by taking the real part of the Fourier transform of the nonlinear
third-order response functions as
\begin{align}
S^{(3)}(t_{3},t_{2},t_{1}&)=\left(\frac{-i}{\hbar}\right)^{3}\times \nonumber\\
&\left\langle \mu(t_{3})[\mu(t_{2}),[\mu(t_{1}),[\mu(0),\rho(-\infty)]]]\right\rangle   ,
\end{align}
where $\hbar$ is the reduced Planck constant, $\mu(t)$ is the dipole operator in the interaction picture,
and $\rho(-\infty)$ is the density matrix at thermal equilibrium.
It can be expanded into four terms and their complex conjugates as
\begin{equation}
\begin{split}
R_{1}(&t_{3},t_{2},t_{1}) \!\! =\!\!\left|\mu_{01}\right|^{4}e^{-i\omega(t+\tau)}\times\\
\!\!&\!\!e^{-g(\tau)-g(T_w)-g(t)+g(\tau+T_w)+g(T_w+t)-g(\tau+T_w+t)},\\
R_{2}(&t_{3},t_{2},t_{1}) \!\! =\!\!\left|\mu_{01}\right|^{4}e^{-i\omega(t-\tau)}\times\\
\!\!&\!\!e^{-g(\tau)+g(T_w)-g(t)-g(\tau+T_w)-g(T_w+t)+g(\tau+T_w+t)},\\
R_{3}(&t_{3},t_{2},t_{1}) \!\! =\!\!\left|\mu_{01}\right|^{4}e^{-i\omega(t-\tau)}\times\\
\!\!&\!\!e^{-g(\tau)+g(T_w)-g(t)-g(\tau+T_w)-g(T_w+t)+g(\tau+T_w+t)},\\
R_{4}(&t_{3},t_{2},t_{1}) \!\! =\!\!\left|\mu_{01}\right|^{4}e^{-i\omega(t+\tau)}\times\\
&e^{-g(\tau)-g(T_w)-g(t)+g(\tau+T_w)+g(T_w+t)-g(\tau+T_w+t)},
\end{split}
\end{equation}
where $t_{1}=\tau$, $t_{2}=\tau+T_w$, and $t_{3}=\tau+T_w+t$ are the
delay times, $\omega$ is the transition frequency between the ground
state $\vert0\rangle$ and the excited state $\vert1\rangle$, $\mu_{01}=\langle0\vert\mu\vert1\rangle$
is the transition dipole between the two states. And $g(t)$ is the
line shape function \cite{Mukamel1999}, which is obtained from the double integration of the TCF as
\begin{align}
g(t)=\frac{1}{2}\int_{0}^{t}\textrm{d}t'\int_{0}^{t'}\textrm{d}t''C(t'')  & .
\end{align}
Generally, the TCF reads \cite{Mukamel1999}
\begin{eqnarray}
C(t)&\equiv &\int \textrm{d}\omega J(\omega)\left[\coth\left (\frac{\beta\omega}{2}\right)\cos(\omega t)-i\sin(\omega t)\right]\nonumber\\
& \equiv &C'(t)+iC''(t),
\end{eqnarray}
where $C'(t)$ and $C''(t)$ are respectively the real and imaginary parts of the correlation function, $J(\omega)$ is the spectral density characterizing the system-bath interaction, $\beta=1/k_BT$ is the inverse temperature with $k_B$ and $T$ being the Boltzman constant and the temperature.The real part and imaginary part of the TCF have corresponding relations. When we have the imaginary part, we also know the real part and vice versa,
\begin{equation}
	C''(t)  =\tan \left(\frac{\beta\hbar}{2}\frac{d}{dt}\right) C'(t),
	\label{equ:ctct}
\end{equation}
which will be proven in Appendix~\ref{App:TCF}. In the approach of the center line slope, two approximations are
made to obtain the TCF, i.e., $C(t)=\delta(t)/T_2+\Delta^2\exp(-t/\tau_d)$ \cite{Kwak2007JCP}. First of all, the homogeneous line width is set to zero, i.e., $1/T_{2}=0$. This
approximation ensures that $C(t)$'s motionally-narrowed component
is no longer present. In addition, under short-time approximation, the line shape function is expanded to the second order of the coherence times, i.e., $\tau$ and $t$, as
\begin{align}
g(t) & =\Delta^{2}\tau_d t+\Delta^{2}\tau^{2}(e^{-t/\tau_d}-1)\approx\frac{\Delta^{2}t^{2}}{2},
\end{align}
where the short-time approximation assumes slow spectral dispersion,
i.e., $\Delta\tau_{d}\gg1$.

Subsequently, the absorptive line shape can be written as
\begin{align}
R^{g}(\omega_{\tau},&T_w,\omega_{t})=\frac{2\pi}{\sqrt{C^{2}(0)-C^{2}(T_w)}}\times\nonumber\\
&\exp\left(-\frac{C(0)(\omega_{t}^{2}+\omega_{\tau}^{2})-2C(T_w)\omega_{\tau}\omega_{t}}{2(C^{2}(0)-C^{2}(T_w))}\right).
\end{align}
For a specific $\omega_{t}$, we can obtain the maximum by calculating
the derivative with respect to $\omega_{\tau}$ as
\begin{eqnarray}
&&\frac{\partial R^{g}(\omega_{\tau},T_w,\omega_{t})}{\partial\omega_{\tau}}|_{\omega_{\tau}=\omega_{\tau}^{\textrm{max}}}\nonumber\\
&=&\frac{-C(0)\omega_{\tau}+C(T_w)\omega_{t}}{C^{2}(0)-C^{2}(T_w)}
\times R^{g}(\omega_{\tau},T_w,\omega_{t})=0.
\end{eqnarray}
The center line is the line connecting the maxima for different $\omega_{t}$'s.
The CLS is the slope of the center line, i.e.,
\begin{equation}
\text{CLS}\omega_{t}(T_w)=\frac{\textrm{d}\omega_{\tau}^{\textrm{max}}(\omega_{t})}{\textrm{d}\omega_{t}}=\frac{C(T_w)}{C(0)}.
\end{equation}
Alternatively, the CLS can be obtained in a similar way as
\begin{align}
\text{CLS}\omega_{\tau}(T_w) & =\frac{\textrm{d}\omega_{t}^{\textrm{max}}(\omega_{\tau})}{\textrm{d}\omega_{\tau}}=\frac{C(T_w)}{C(0)},
\end{align}
where the maximum is determined by
\begin{eqnarray}
&&\frac{\partial R^{g}(\omega_{\tau},T_w,\omega_{t})}{\partial\omega_{t}}|_{\omega_{t}
=\omega_{t}^{\textrm{max}}}\nonumber\\
&=&\frac{-C(0)\omega_{t}+C(T_w)\omega_{\tau}}{C^{2}(0)-C^{2}(T_w)}
\times R^{g}(\omega_{\tau},T_w,\omega_{t})=0.
\end{eqnarray}

Note that the CLS method is based on the response function, which
treats all excitation pulses as delta pulses. Furthermore, several
approximations are employed, such as the short-time approximation, omitting
the homogeneous term and assuming the TCF to be real. Therefore,
it is quite natural to question the validity of the CLS approach under a more realistic condition, e.g. employing it in the 2D electronic spectroscopy in the visible-frequency domain at low temperatures 

\section{CLS at Low Temperature}
\label{Sec:SimulatedSpectrum}

In the original CLS approach, a real function is assumed for the TCF. It can be viewed as the high-temperature limit of the natural photosynthetic complexes. In this section, we will test the reliability of the CLS approach at a low temperature, i.e., a complex function \cite{meier1999jcp} $C(t)=a(t)-ib(t)$, where
\begin{align}\label{tcfform}
a(t) & =\sum_k\alpha_ke^{-\gamma_k t},\\
b(t) & =\sum_k\alpha_ke^{-\gamma_k t}.
\end{align}
  Tab.~\ref{tab:para1} shows the parameters of the complex function when the temperature is 298~K. The real and imaginary parts of the TCF should satisfy the corresponding relationship Eq.~(\ref{equ:ctct}).We also considered the case of temperature 77~K, with parameters given in Tab.~\ref{tab:para2}.

\begin{table}
	\centering
	
	\begin{tabular}{cccc}
		\hline
		TCF & $\alpha_k$($\times10^{2}$fs$^{-2}$) & $\text{Re}(\gamma_k)$($\times10^{-3}$fs$^{-1}$) & $\text{Im}(\gamma_k)$($\times10^{-3}$fs$^{-1}$) \\
		\hline
		$a(t)$ & $65.24$ & $9.183$  & $0$  \\
		$a(t)$ & $9.001$ & $2.336$  & $0$  \\
		$a(t)$ & $-10.56$ & $28.62$  & $19.17$  \\
		$a(t)$ & $-10.56$ & $28.62$  & $-19.17$  \\
		$b(t)$ & $-0.9246$ & $6.374$ & $-5.870$ \\
		$b(t)$ & $-0.9246$ & $6.374$ & $5.870$  \\
		$b(t)$ & $2.655$ & $20.76$ & $-20.48$   \\
		$b(t)$ & $2.655$ & $20.76$ & $20.48$ \\
		$b(t)$ & $-3.717$ & $5.360$ & $0$ \\
		\hline
	\end{tabular}
	\caption{Parameters of TCF at $T$=298~K}
	\label{tab:para1}
\end{table}

\begin{table}
	\centering
	
	\begin{tabular}{cccc}
		\hline
		TCF & $\alpha_k$($\times10^{3}$fs$^{-2}$) & $\text{Re}(\gamma_k)$($\times10^{-3}$fs$^{-1}$) & $\text{Im}(\gamma_k)$($\times10^{-5}$fs$^{-1}$) \\
		\hline
		$a(t)$ & $11.82$ & $17.51$  & $-2.777$  \\
		$a(t)$ & $11.82$ & $17.51$  & $2.777$  \\
		$a(t)$ & $-22.64$ & $18.40$  & $0$  \\
		$a(t)$ & $0.4852$ & $3.331$  & $0$  \\
     	$b(t)$ & $-2.879$ & $23.77$ & $-3.043$ \\
    	$b(t)$ & $-2.879$ & $23.77$ & $3.043$ \\
    	$b(t)$ & $3.022$ & $24.08$ & $-781.9$ \\
    	$b(t)$ & $3.022$ & $24.08$ & $781.9$ \\
    	$b(t)$ & $-0.2854$ & $5.595$ & $0$ \\
		\hline
	\end{tabular}
	\caption{Parameters of TCF at $T$=77~K}
	\label{tab:para2}
\end{table}

\begin{figure}
	\begin{centering}
		\includegraphics[bb=0 0 185 145,width=7cm]{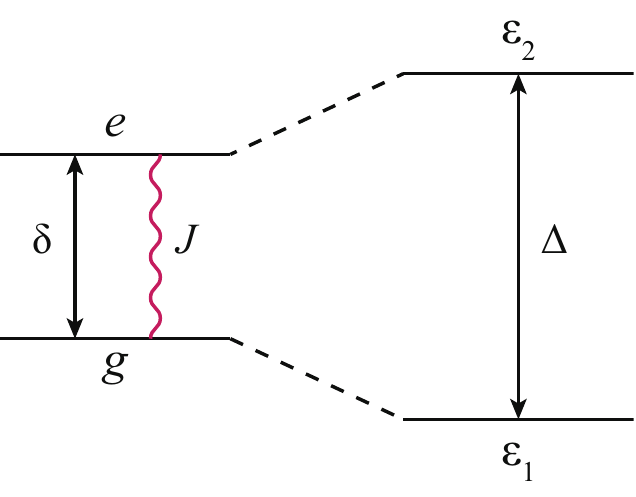}
	\end{centering}
	\caption{ The two-level atom with coupling strength $J$, where $\delta\equiv\epsilon_2-\epsilon_1=400~\textrm{cm}^{-1}$ is the
		energy gap between $\left|e\right\rangle$ and $\left|g\right\rangle$. When there is resonant coupling $J=300~\text{cm}^{-1}$ between $\left|e\right\rangle$ and $\left|g\right\rangle$, the energy gap has been widened as $\Delta=\sqrt{\delta^2+J^2}$. }
\label{fig:scheme}
\end{figure}

\begin{figure*}
	\begin{centering}
		\includegraphics[width=18cm]{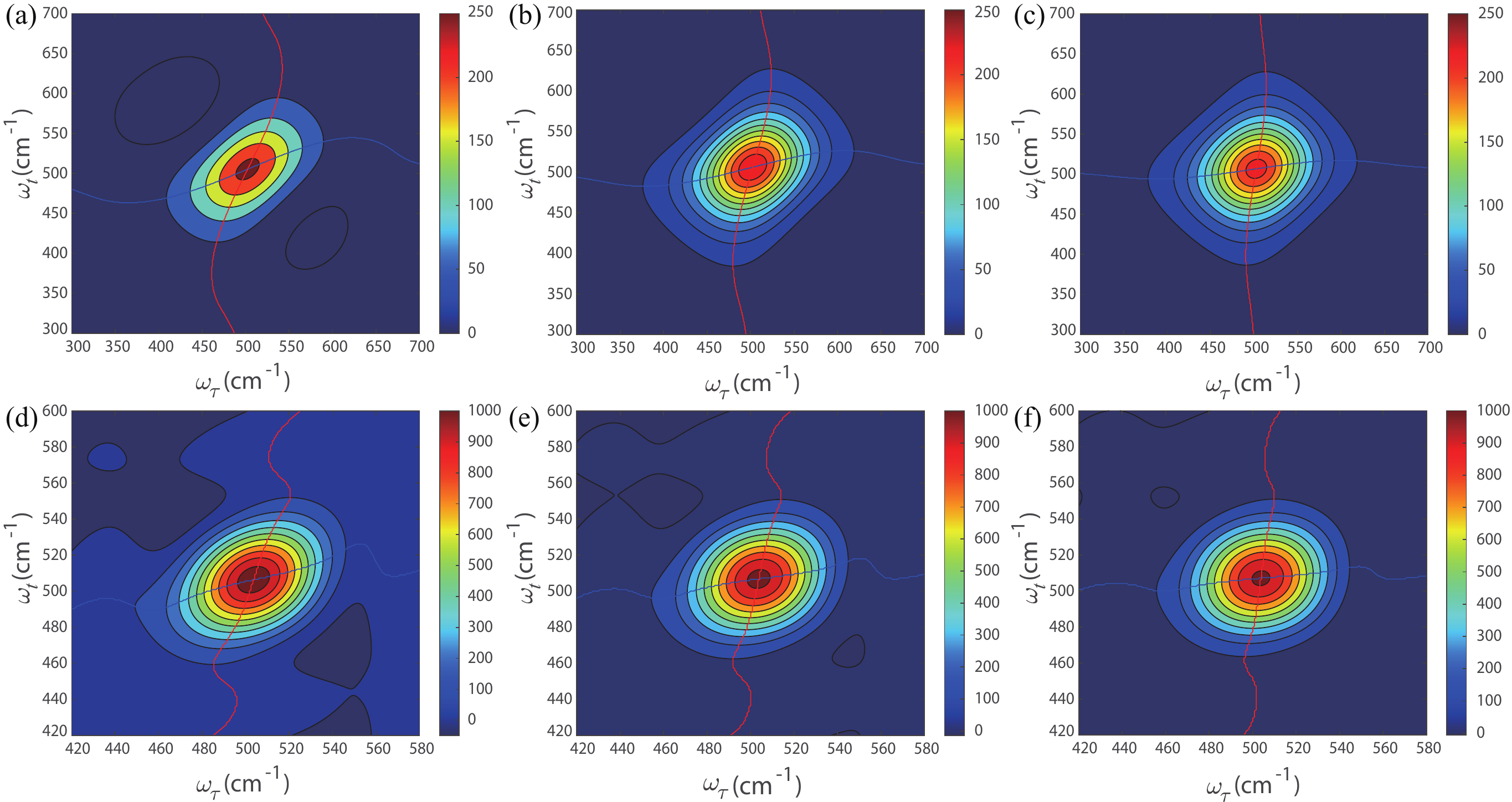}
		\end{centering}
	\caption{2DES spectra simulated from the response function at the waiting time (a,d) $T_w= 10$~fs; (b,e) 100~fs; and (c,f) 200~fs. The red (blue) solid line is the center line for $\omega_{t}$ ($\omega_{\tau}$). The sub-figures (a-c) are simulated with $T=298$~K, while (d-f) with $T=77$~K.}
\label{fig:2DfiniteTemp}
\end{figure*}

In Fig.~\ref{fig:2DfiniteTemp}, we show the 2DES spectra generated from the response function for the two-level atom with coupling strength $J$ as schematically depicted in Fig.~\ref{fig:scheme}. The red (blue) solid lines in the diagram are the center lines for $\omega_{t}$ ($\omega_{\tau}$). In the neighbourhood of the peak, these two lines are very close to a straight line, and their slopes are the CLS. In order to obtain the behavior of the TCF in the time domain, we calculate the 2DES for a series of waiting times
$T_w$. With the increase of $T_w$, the shape of the peak changes from ellipse to circle due to  the combination of homogeneous dephasing and spectral diffusion \cite{Kwak2008JCP}. Moreover, in the long-time limit, the two center lines tend to be parallel to the two coordinate axes respectively. According to the CLS approach, since the slopes of the two lines gradually approach 0, the TCF eventually vanishes.

\begin{figure}
\includegraphics[width=8cm]{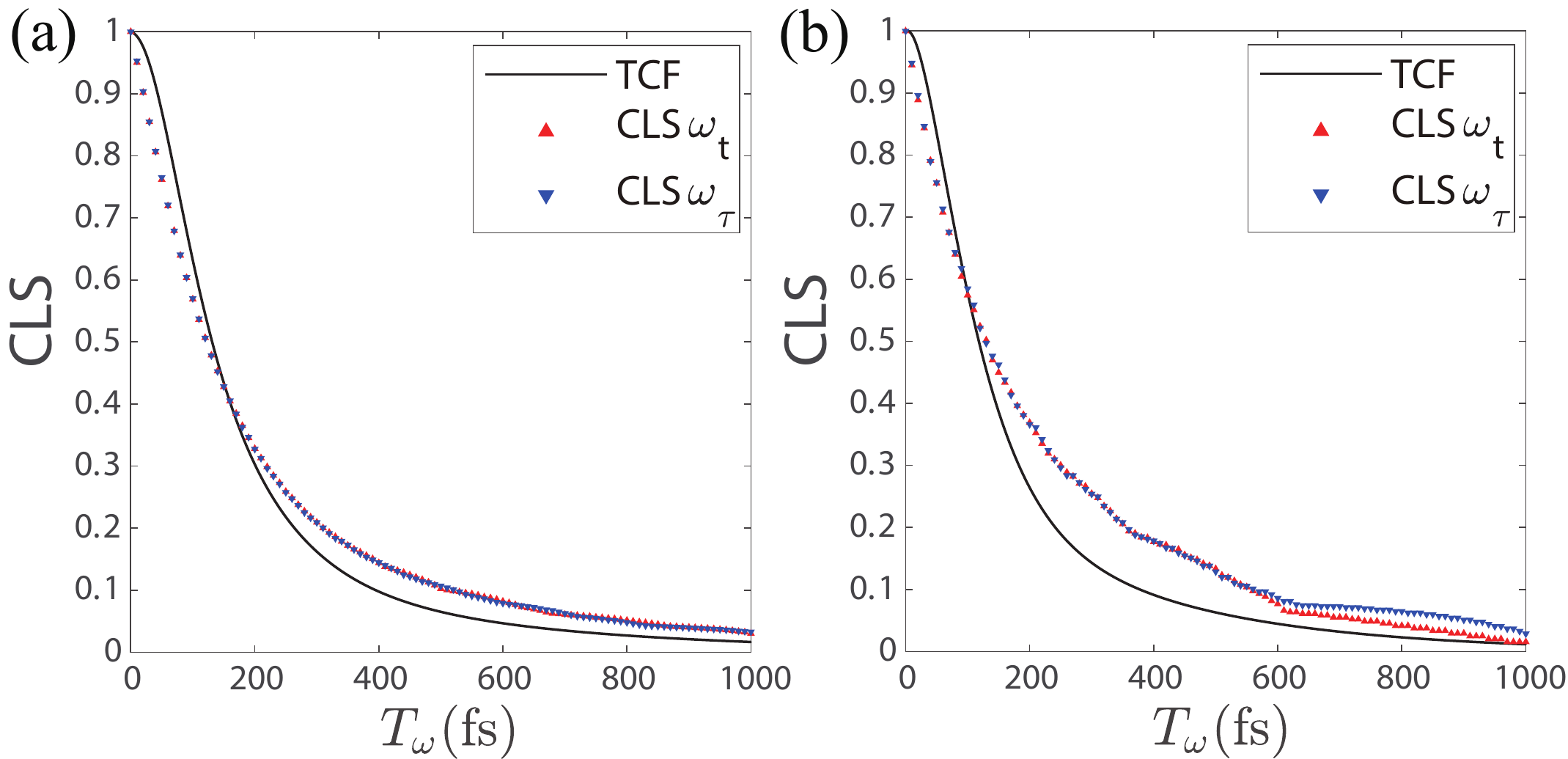}
\caption{The dependence of the CLS on the waiting time $T_w$, where CLS is simulated with (a) $T=298$~K, (b) $T=77$~K. The black line is the real part of the TCF.}
\label{fig:CLSvsTemp}
\end{figure}

Notice that when it is much away from the peak, the center line significantly deviates from a straight line. In order to effectively obtain the TCF, we restrict the center line to the full width at the half maximum (FWHM) of the peak and numerically fit it to obtain the CLS. The relation between the CLS and the waiting time $T_w$ is shown in Fig.~\ref{fig:CLSvsTemp}, where the CLS is normalized by its value at $T_w=0$~fs.
Obviously, the CLS$\omega_{t}$ and CLS$\omega_{\tau}$ at the same temperature are almost the same, and the behavior of the CLS in the time domain is similar to that of the TCF.


\section{The effect of static disorder on the center line slope}
\label{Sec:StaticDisorder}

Hereafter, we calculate the 2DES with the energy detuning characterized by a
Gaussian distribution $\delta$. Assuming there is electronic coupling $J$ between the two levels, the probability of the energy gap $\Delta$ between the two eigenstates is derived in Appendix~\ref{App:Distribution}. In the paper, we use the 128-point Gauss--Hermite quadrature method \cite{Abramowitz1964} to calculate the spectra for each waiting time $T_w$.

For static disorder in the range 10-200~$\textrm{cm}^{-1}$, we calculate the 2DES and obtain the corresponding CLS. We find that the 2DES is modified mainly in three aspects due to the static disorder, i.e., the shape of the spectral peak, the steady-state value of the CLS in the long-time limit and the decay rate $\gamma$ of the CLS.

\subsection{The shape of the spectral peak}

\begin{figure*}
\begin{centering}
\includegraphics[width=18cm]{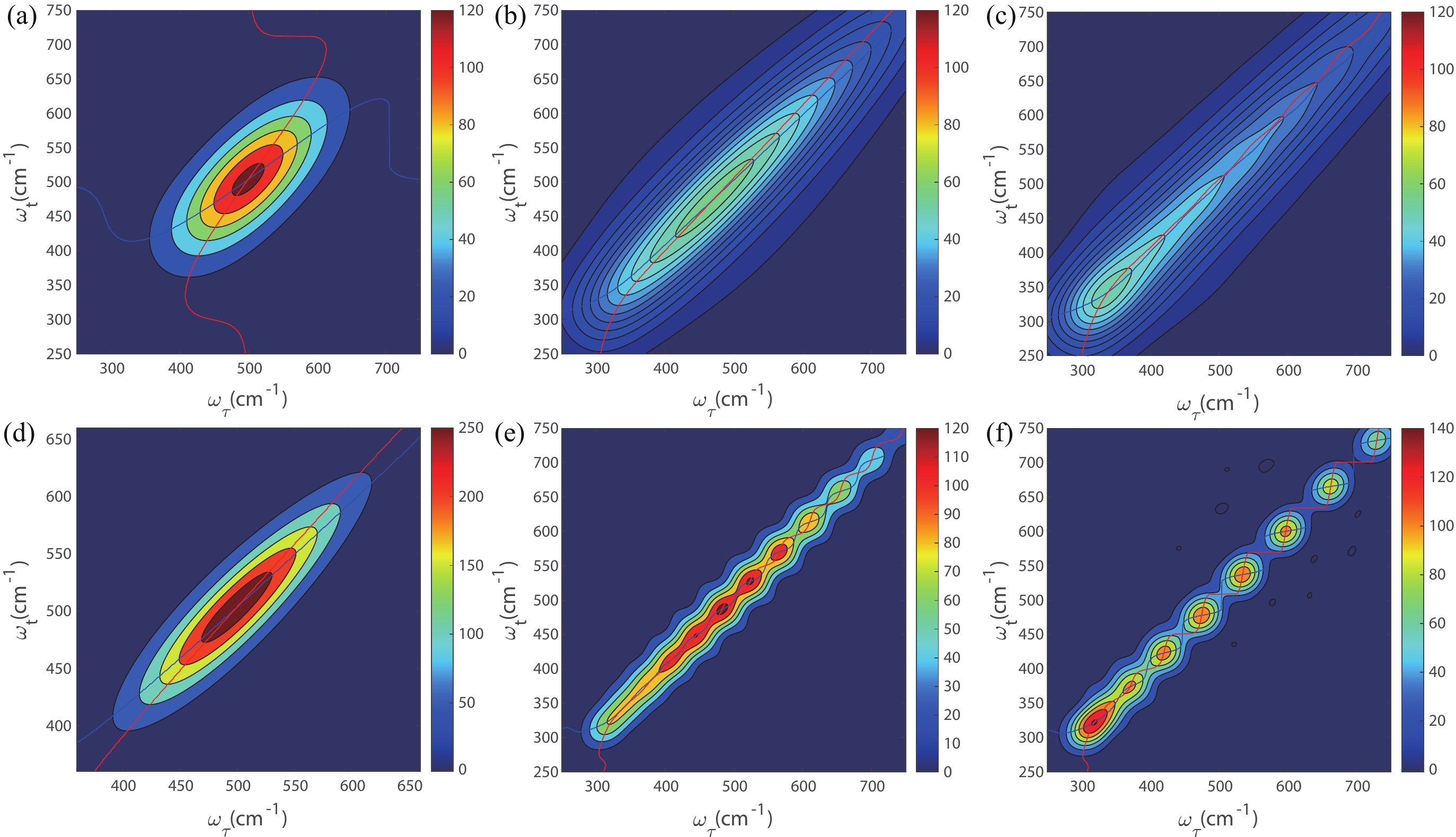}
\end{centering}
\caption{2DES spectra simulated with the static disorder (a,d) $\sigma=50~\textrm{cm}^{-1}$, (b,e) $\sigma=130~\textrm{cm}^{-1}$, and (c,f) $\sigma=190~\textrm{cm}^{-1}$. The figures (a-c) are simulated with $T=298$~K, while (d-f) with $T=77$~K.}
\label{fig:2DfiniteTemp2}
\end{figure*}

In Fig.~\ref{fig:2DfiniteTemp2}, we show the 2DES of $T_w$=100~fs for three typical static disorders. Compared with Fig.~\ref{fig:2DfiniteTemp}(b), the peak has been stretched along the diagonal line as a result of the static disorder.
With the increase of the static disorder, the stretching effect becomes
more obvious. Notice that when the static disorder is greater
than 190~cm$^{-1}$, a series of smaller peaks with lower height emerge
in the diagonal line and the center line appears wavy, which is significantly
different from a straight line. This effect will significantly affect the acquisition of the CLS, making it difficult to obtain the TCF. It is worth noting that the impact of static disorder is greater at the case that $T$=77~K compared to $T$=298~K. For example, at $\sigma=130~\textrm{cm}^{-1}$, the spectrum at $T$=298~K is affected by the static disorder causing the peak shape to be stretched along the diagonal direction. However, the stretching effect at  $T$=77~K is more intense, splitting into multiple smaller peaks along the diagonal direction. A similar effect does not appear in the 298~K graph until $\sigma$ increases to 190~$\textrm{cm}^{-1}$.

\subsection{The steady-state value of the CLS}

Another effect of the static disorder is reflected in the steady-state value of the CLS in long-time limit. In the absence of the static disorder, the CLS goes to 0, which is the same as the behavior of the TCF. However, we find that when the static disorder is introduced, the CLS at the steady state no longer vanishes. This steady-state value increases when the amplitude of static disorder is enlarged.

\begin{figure}
	\begin{centering}
		\includegraphics[width=8.5cm]{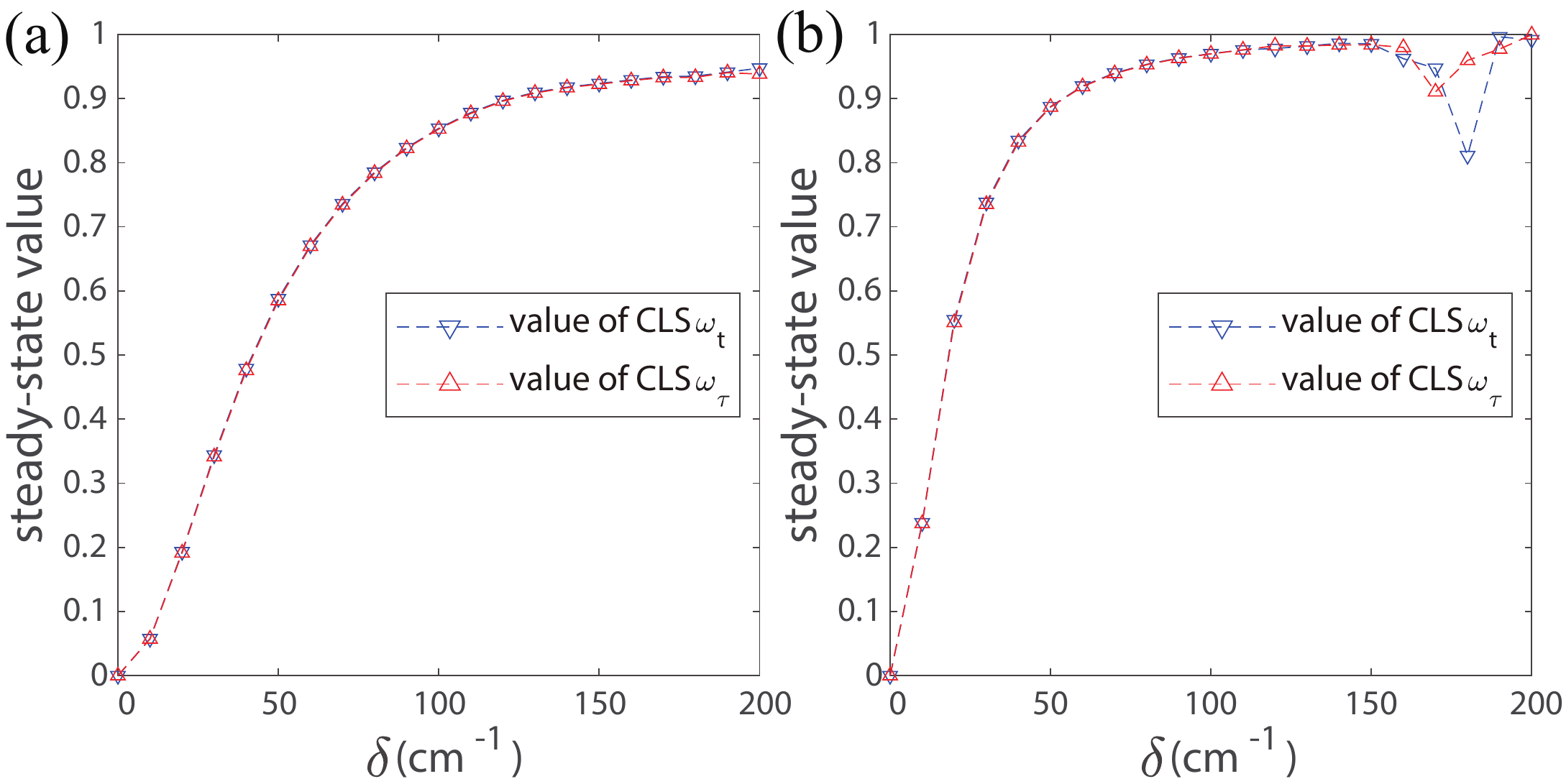}
	\end{centering}
	\caption{The steady-state value of CLS at (a) $T=298$~K, (b) $T=77$~K varies with the amplitude of static disorder. The red (blue) line represents the change of CLS$\omega_\tau$ (CLS$\omega_t$). }
	\label{fig:steady}
\end{figure}

Figure~\ref{fig:steady} shows the change of the steady-state value with the static disorder. We find that there is a monotonic relationship between the steady-state value of the CLS and the amplitude of static disorder. As the latter increases, the steady-state value of CLS increases rapidly. When $\delta$ raises to 50~cm$^{-1}$, the CLS ultimately decays only to $60\%$ of its initial value. At a temperature of 298~K, the steady-state values of CLS$\omega_{\tau}$ and CLS$\omega_{t}$ overlap with each other in whole parameter regime, as shown in Fig.~\ref{fig:steady}(a). However, at a lower temperature of 77~K as shown in Fig.~\ref{fig:steady}(b), When $\sigma$ exceeds 160~$=\textrm{cm}^{-1}$, an abnormal fluctuation in the steady-state values of CLS$\omega_{\tau}$ and CLS$\omega_{t}$ appears at 77~K. We speculate that this is due to the excessive impact of static disorder on the peak shape, causing the CLS method to fail to extract information correctly. In comparison to the situation at 77~K, the impact of static disorder is smaller at 298~K and a similar fluctuation does not appear in Fig.~\ref{fig:steady}(a).

\subsection{The decay rate of the CLS}

The static disorder also affects the decay rate of the CLS. For a group of spectra with different $T$'s, we perform a single exponential fitting, i.e., in the form of $ae^{-\gamma t}$, to obtain the decay rate~$\gamma$ of the CLS, which is very crucial in studying open quantum dynamics. The actual TCF, as shown in Eq.~(\ref{tcfform}), is the sum of multiple exponential equations, but fitting data with multiple exponentials results in many sets of fitting results due to too many parameters, which is not helpful in obtaining information about the interaction between the system and the environment, as von Neumann said\cite{Mayer2010AJP,dyson2004Nature}., "With four parameters I can fit an elephant, and with five I can make him wiggle his trunk." Therefore, we use a single exponential for fitting, which may not perfectly match the data, but can capture the main decay component of the TCF and provide the dominant interaction component between the system and the environment.

\begin{figure}
	\begin{centering}
		\includegraphics[width=8.5cm]{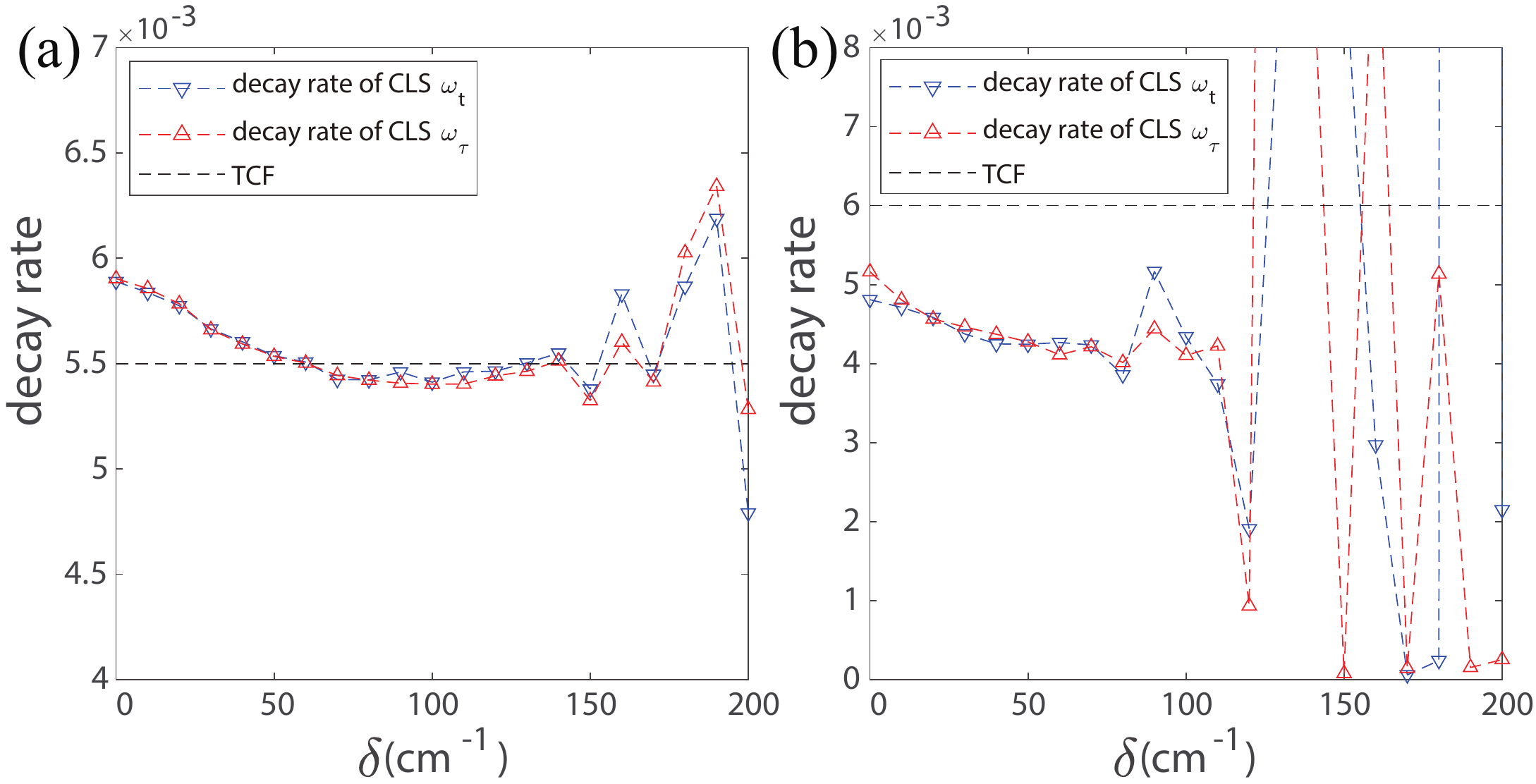}
	\end{centering}
	\caption{The decay rate of CLS at (a) $T=298$~K, (b) $T=77$~K varies with the amplitude of static disorder. The red (blue) line represents the change of CLS$\omega_\tau$ (CLS$\omega_t$). The black dotted line is the decay rate of the TCF.}
	\label{fig:decayrate}
\end{figure}

Figure~\ref{fig:decayrate} shows the decay rate of the CLS under different static disorders, where the black dotted line is the decay rate of the TCF. We find that at both temperatures, i.e., 298~K and 77~K, the decay rate of the CLS exhibits a similar relationship with the amplitude of static disorder. When the static disorder emerges, as it increases, the decay rate of the CLS decreases, causing the decay to slow down. When the decay rate decreases to a certain value, it stops decreasing and remains as a constant regardless of the increasing static disorder. In Fig~\ref{fig:decayrate}(a), as the static disorder is larger than about 130~cm$^{-1}$, the decay rate begins to fluctuate. When $\sigma$ increases to nearly 200~cm$^{-1}$, the CLS theory fails to extracting the TCF, because additional peak emerge due to the large static disorder, as shown in Fig.~\ref{fig:2DfiniteTemp2}(c) and (f). In Fig~\ref{fig:decayrate}(b), a highly obvious fluctuation appears when the static disorder is greater than 80~cm$^{-1}$, which is more affected than in the case of 298~K.

\begin{table}[htbp]
	\centering
	\begin{tabular}{ccc}
		\hline
		$\sigma$(cm$^{-2}$) & $\Delta\gamma_t$($\times 10^{-4}$fs$^{-1}$) & $\Delta\gamma_\tau$($\times 10^{-4}$fs$^{-1}$) \\
		\hline
		0 & 6.55  & 3.02  \\
		10 & -3.735  & -3.921  \\
		20 & -3.109  & -3.202  \\
		30 & -2.009  & -1.975  \\
		40 & -1.394  & -1.294  \\
		50 & -0.7414$$ & -0.7095  \\
		60 & -0.4401  & -0.3938  \\
		70 & 0.3799  & 0.1916  \\
		80 & 0.3984  & 0.4130  \\
		90 & 0.02218  & 0.5534  \\
		100 & 0.4969  & 0.5941  \\
		110 & 0.01419  & 0.5630  \\
		120 & -0.02542  & 0.1967  \\
		130 & -0.4144  & -0.03379  \\
		140 & -0.8844  & -0.4954  \\
		150 & 0.8333  & 1.369  \\
		160 & -3.669  & -1.361  \\
		170 & 0.06429  & 0.4112  \\
		180 & -4.025  & -5.632  \\
	\end{tabular}
	\caption{Correction of decay rate at 298~K}
	\label{tab:table2}
\end{table}

\begin{table}[htbp]
	\centering
	\begin{tabular}{ccc}
		\hline
		$\sigma$(cm$^{-2}$) & $\Delta\gamma_t$($\times 10^{-3}$fs$^{-1}$) & $\Delta\gamma_\tau$($\times 10^{-3}$fs$^{-1}$) \\
		\hline
		0     & 1.168  & 0.815   \\
		10    & 1.2659 & 1.1718  \\
		20    & 1.3909 & 1.4112  \\
		30    & 1.6023 & 1.5123  \\
		40    & 1.7258 & 1.6027  \\
		50    & 1.7236 & 1.7016  \\
		60    & 1.7124 & 1.8517  \\
		70    & 1.7357 & 1.7810  \\
		80    & 2.1244 & 1.9769  \\
		90    & 0.7453 & 1.5994  \\
		100   & 1.5299 & 1.8817  \\
	\end{tabular}
	\caption{Correction of decay rate at 77~K}
	\label{tab:table3}
\end{table}

\section{Conclusion}
\label{Sec:Dicussion}

To summarize the main findings of the paper, it is shown that the static disorder has a significant impact on the reliability of the CLS method for obtaining the TCF from 2DES. Specifically, we find that the presence of the static disorder leads to an increase in the steady-state value of the CLS and an inaccurate decay rate of the CLS, as compared to the TCF.

We show that there is a monotonic relation between the steady-state value of the CLS and the amplitude of the static disorder. As the static disorder increases, the steady-state value of the CLS also increases, as shown in  Fig.~\ref{fig:steady}. According to simulated data at 77~K and 298~K, the relation between the steady-state value and the static disorder is very similar. Due to this monotonic relation, it is possible to obtain the amplitude of the static disorder of the system through the CLS approach.

We also find that the amplitude of the static disorder affects the decay rate of the CLS. When using the CLS method to extract the TCF in the laboratory, the presence of the static disorder can lead to errors in the decay rate, especially in the case with large static disorder. Since we have the correspondent error of the decay rate of the CLS and the amplitude of the static disorder, cf. Fig.~\ref{fig:decayrate}, we propose correcting the decay rate of the CLS to achieve the decay rate of the TCF.

Our work can provide a correction for the application of CLS method in actual experiments, where we define the correction $\Delta\gamma =\gamma_{\text{TCF}}-\gamma_{\text{CLS}}$. $\gamma_{\text{TCF}}$ and $\gamma_{\text{CLS}}$ correspond to the decay rates of true TCF and CLS respectively. In practice, by summing up the obtained decay rate $\gamma$ with corresponding correction $\Delta\gamma$ according to static disorder $\sigma$, a result which is much closer to that of true TCF can be attained. The corrections for different static disorder are presented in Tab.~\ref{tab:table2} and Tab.~\ref{tab:table3}.

\begin{acknowledgments}
This work is supported by Beijing Natural Science Foundation under Grant No.~1202017 and the National Natural Science Foundation of China under Grant Nos.~11674033, 11505007, and Beijing Normal University under Grant No.~2022129.

\end{acknowledgments}

\appendix

\section{Distribution with Static Disorder}
\label{App:Distribution}

The spectra in Sec.~\ref{Sec:SimulatedSpectrum} are computed
without the static disorder. In this appendix, we consider the effect
of the static disorder on the simulation, which is closer to the actual experimental spectrum.

For the two-level system, we consider the static disorder on both levels, which can be described by a Gaussian distribution function
with mean $\omega_{e}^{0}$ ($\omega_{g}^{0}$) and standard deviation
$\sigma_{D}$ ($\sigma_{D}$) for the excited (ground) state $\left|e\right\rangle $
($\left|g\right\rangle $), i.e.,
\begin{align}
\begin{split}
P(\omega_{e}) & =\frac{1}{\sqrt{2\pi}\sigma_{D}}\exp\left(-\frac{(\omega_{e}-\omega_{e}^{0})^{2}}{2\sigma_{D}^{2}}\right),\\
P(\omega_{g}) & =\frac{1}{\sqrt{2\pi}\sigma_{D}}\exp\left(-\frac{(\omega_{g}-\omega_{g}^{0})^{2}}{2\sigma_{D}^{2}}\right).
\end{split}
\end{align}
The energy gap $\omega_{eg}$ between $\left|e\right\rangle $ and
$\left|g\right\rangle $ should also be subject to a Gaussian distribution
\begin{equation}
P(\omega_{eg})=\frac{1}{\sqrt{2\pi}\sigma'_{D}}\exp\left(-\frac{(\omega_{eg}-\omega_{eg}^{0})^{2}}{2(\sigma'_{D})^{2}}\right),\label{eq:Pw}
\end{equation}
where $\omega_{eg}^{0}=\omega_{e}^{0}-\omega_{g}^{0}$ is the mean,
and the standard deviation is modified as $\sigma'_{D}=\sqrt{2}\sigma_{D}$.

However, when there is interaction $J$ between the two levels, the
energy gap reads
\begin{align}
\Delta & =\sqrt{\delta^{2}+J^{2}}
\end{align}
where $\delta$ satisfies the Gaussian distribution with mean $\omega_{eg}^{0}$ and standard deviation $\sigma'_{D}$, i.e.,
\begin{align}
P(\delta)= & \frac{1}{\sqrt{2\pi}\sigma_{D}^{\prime}}\exp[-\frac{(\delta-\omega_{eg}^{0})^{2}}{2\sigma_{D}^{\prime2}}].
\end{align}
By the normalization condition, we have
\begin{eqnarray}
\int_{-\infty}^{+\infty}P(\delta)d\delta&= & \int_{-\infty}^{+\infty}\frac{1}{\sqrt{4\pi}\sigma_{D}}\exp\left(-\frac{(\delta-\omega_{eg}^{0})^{2}}{4\sigma_{D}^{2}}\right)d\delta\nonumber \\
&= & \int_{0}^{+\infty}\frac{1}{\sqrt{4\pi}\sigma_{D}}\exp\left(-\frac{(\delta-\omega_{eg}^{0})^{2}}{4\sigma_{D}^{2}}\right)d\delta\nonumber\\
&&+\frac{1}{\sqrt{4\pi}\sigma_{D}}\exp\left(-\frac{(\delta+\omega_{eg}^{0})^{2}}{4\sigma_{D}^{2}}\right)d\delta\nonumber\\
&= & 1.\label{eq:Normalization}
\end{eqnarray}
Since
\begin{align}
\delta=\sqrt{\Delta^{2}-J^{2}},\label{eq:delta}
\end{align}
we have
\begin{align}
d\delta=\frac{\Delta}{\sqrt{\Delta^{2}-J^{2}}}d\Delta.\label{eq:Ddelta}
\end{align}
By substituting Eqs.~(\ref{eq:delta}) and~(\ref{eq:Ddelta}) into
Eq.~(\ref{eq:Normalization}), the probability distribution of the energy gap with interaction is given as \cite{Dong2014jpcb}
\begin{align}
P(\Delta) & =\frac{\exp\left(-\frac{(\sqrt{\Delta^{2}-J^{2}}-\omega_{eg}^{0})^{2}}{4\sigma_{D}^{2}}\right)+\exp\left(-\frac{(\sqrt{\Delta^{2}-J^{2}}+\omega_{eg}^{0})^{2}}{4\sigma_{D}^{2}}\right)}{\sqrt{4\pi}\sigma_{D}\Delta^{-1}\sqrt{\Delta^{2}-J^{2}}}.
\end{align}

\section{Relation between Real and Imaginary Parts of TCF}
\label{App:TCF}

Generally, the TCF $C\left(t\right)$ is complex, and we divide
it into its real and imaginary parts as $C\left(t\right)=C'\left(t\right)+iC''\left(t\right).$
In the frequency domain, the Fourier transform of the TCF
reads
\begin{align}
\tilde{C}(\omega) & =2{\rm Re}\int_{0}^{\infty}dt\:e^{i\omega t}C\left(t\right)\nonumber\\
 & =2{\rm Re}\int_{0}^{\infty}dt(\cos\omega t+i\sin\omega t)\left(C'\left(t\right)+iC''\left(t\right)\right)\nonumber\\
 & =2\int_{0}^{\infty}dt\left(\cos\omega tC'\left(t\right)-\sin\omega tC''\left(t\right)\right)\nonumber\\
 & =2\int_{0}^{\infty}dt\cos\omega tC'\left(t\right)-2\int_{0}^{\infty}dt\sin\omega tC''\left(t\right)\nonumber\\
 & =\tilde{C}'(\omega)-\tilde{C}''(\omega),
\label{eq:Cw}
\end{align}
where the real and imaginary parts of $\tilde{C}'(\omega)$ are respectively

\begin{align}
\begin{split}
\tilde{C}'(\omega) & =2\int_{0}^{\infty}dt\cos\left(\omega t\right)C'\left(t\right),\\
\tilde{C}''(\omega) & =2\int_{0}^{\infty}dt\sin\left(\omega t\right)C''\left(t\right).
\end{split}
\end{align}
By inverse Fourier transform, we have

\begin{equation}\label{eq:Ct}
	\begin{split}
		C'\left(t\right)&=\frac{1}{\pi}\int_{0}^{\infty}d\omega\cos\left(\omega t\right)\tilde{C}'(\omega),\\
		C''\left(t\right)&=\frac{1}{\pi}\int_{0}^{\infty}d\omega\sin\left(\omega t\right)\tilde{C}''(\omega).
	\end{split}
\end{equation}

where
\begin{equation}\label{eq:OddEven}
	\begin{split}
		\tilde{C}'(\omega)&=\tilde{C}'(-\omega),\\
		\tilde{C}''(\omega)&=-\tilde{C}''(-\omega).
	\end{split}
\end{equation}


$\tilde{C}\left(\omega\right)$ satisfies the detailed-balance condition \cite{Mukamel1999}
\begin{eqnarray}
\tilde{C}(-\omega)=e^{-\beta\hbar\omega}\tilde{C}(\omega).
\label{eq:DeBal}
\end{eqnarray}
By substituting Eq.~(\ref{eq:OddEven}) and Eq.~(\ref{eq:DeBal}) into Eq.~(\ref{eq:Cw}), we can obtain
\begin{equation}
\begin{split}
\tilde{C}'(\omega)=\frac{e^{-\beta\hbar\omega}+1}{2}\tilde{C}(\omega),\\
\tilde{C}''(\omega)=\frac{e^{-\beta\hbar\omega}-1}{2}\tilde{C}(\omega).
\end{split}
\end{equation}
Then, we have
\begin{eqnarray}
\tilde{C}''(\omega) & =&\frac{e^{-\beta\hbar\omega}-1}{e^{-\beta\hbar\omega}+1}\tilde{C}'(\omega)\nonumber\\
&=&\frac{e^{-\frac{\beta\hbar\omega}{2}}-e^{\frac{\beta\hbar\omega}{2}}}{e^{-\frac{\beta\hbar\omega}{2}}+e^{\frac{\beta\hbar\omega}{2}}}\tilde{C}'(\omega)\nonumber\\
&=&-\tanh\left(\frac{\beta\hbar\omega}{2}\right)\tilde{C}'(\omega),
\end{eqnarray}
or equivalently
\begin{equation}
\tilde{C}'(\omega)=-\coth\left(\frac{\beta\hbar\omega}{2}\right)\tilde{C}''(\omega).\label{eq:Frela}
\end{equation}
By substituting Eq.~(\ref{eq:Frela}) into to Eq.~(\ref{eq:Ct}), we obtain

\begin{equation}
	\begin{split}
		C'\left(t\right)&=-\int_{0}^{\infty}d\omega\frac{\cos\left(\omega t\right)}{\pi}\coth\left(\frac{\beta\hbar\omega}{2}\right)\tilde{C}''(\omega),\\
	C''\left(t\right)&=\int_{0}^{\infty}d\omega\frac{\sin\left(\omega t\right)}{\pi}\tilde{C}''(\omega).
	\end{split}
\end{equation}

\begin{widetext}
Since

\begin{align}
\left(\frac{\beta\hbar}{2}\frac{d}{dt}\right)^{2n-1}C'\left(t\right) & =\left(\frac{\beta\hbar}{2}\frac{d}{dt}\right)^{2n-1}\left(-\frac{1}{\pi}\int_{0}^{\infty}d\omega\tilde{C}''(\omega)\coth\left(\frac{\beta\hbar\omega}{2}\right)\cos\omega t\right)\nonumber\\
 & =\left(\frac{\beta\hbar}{2}\right)^{2n-1}\left(\frac{d}{dt}\right)^{2n-1}\left(-\frac{1}{\pi}\int_{0}^{\infty}d\omega\tilde{C}''(\omega)\coth\left(\frac{\beta\hbar\omega}{2}\right)\cos\omega t\right)\nonumber\\
 & =\left(\frac{\beta\hbar}{2}\right)^{2n-1}\left(-\frac{1}{\pi}\int_{0}^{\infty}d\omega\tilde{C}''\left(\omega\right)\coth\left(\frac{\beta\hbar\omega}{2}\right)\left(\frac{d}{dt}\right)^{2n-1}\cos\omega t\right)\nonumber\\
 & =\left(\frac{\beta\hbar}{2}\right)^{2n-1}\left(-\frac{1}{\pi}\int_{0}^{\infty}d\omega\tilde{C}''\left(\omega\right)\coth\left(\frac{\beta\hbar\omega}{2}\right)\left(-1\right)^{n}\omega^{2n-1}\sin\omega t\right)
\end{align}
we have
\begin{eqnarray}
&&\tan\left(\frac{\beta\hbar}{2}\frac{d}{dt}\right)C'\left(t\right)\nonumber\\ & =&\sum_{n=1}^{\infty}\frac{\left(-1\right)^{n-1}2^{2n}\left(2^{2n}-1\right)B_{2n}}{\left(2n\right)!}\left(\frac{\beta\hbar}{2}\right)^{2n-1}\left[-\frac{1}{\pi}\int_{0}^{\infty}d\omega\tilde{C}''\left(\omega\right)\coth\left(\frac{\beta\hbar\omega}{2}\right)\left(-1\right)^{n}\omega^{2n-1}\sin\omega t\right]\nonumber\\
 & =&-\frac{1}{\pi}\int_{0}^{\infty}d\omega\sum_{n=1}^{\infty}\frac{\left(-1\right)^{n-1}2^{2n}\left(2^{2n}-1\right)B_{2n}}{\left(2n\right)!}\left(\frac{\beta\hbar\omega}{2}\right)^{2n-1}\left(-1\right)^{n}\tilde{C}''(\omega)\coth\left(\frac{\beta\hbar\omega}{2}\right)\sin\omega t\nonumber\\
 & =&-\frac{1}{\pi}\int_{0}^{\infty}d\omega\sum_{n=1}^{\infty}\frac{\left(-1\right)^{n-1}2^{2n}\left(2^{2n}-1\right)B_{2n}}{\left(2n\right)!}\left(\frac{\beta\hbar\omega}{2}\right)^{2n-1}\left(i\right)^{2n-1}i\tilde{C}''(\omega)\coth\left(\frac{\beta\hbar\omega}{2}\right)\sin\omega t\nonumber\\
 & =&-\frac{1}{\pi}\int_{0}^{\infty}d\omega\sum_{n=1}^{\infty}\frac{\left(-1\right)^{n-1}2^{2n}\left(2^{2n}-1\right)B_{2n}}{\left(2n\right)!}\left(i\frac{\beta\hbar\omega}{2}\right)^{2n-1}i\tilde{C}''(\omega)\coth\left(\frac{\beta\hbar\omega}{2}\right)\sin\omega t\nonumber\\
 & =&-\frac{1}{\pi}\int_{0}^{\infty}d\omega\tan\left(i\frac{\beta\hbar\omega}{2}\right)i\tilde{C}''(\omega)\coth\left(\frac{\beta\hbar\omega}{2}\right)\sin\omega t\nonumber\\
 & =&-\frac{1}{\pi}\int_{0}^{\infty}d\omega\frac{\sin\left(i\frac{\beta\hbar\omega}{2}\right)}{\cos\left(i\frac{\beta\hbar\omega}{2}\right)}i\tilde{C}''(\omega)\coth\left(\frac{\beta\hbar\omega}{2}\right)\sin\omega t\nonumber\\
 & =&-\frac{1}{\pi}\int_{0}^{\infty}d\omega\frac{\exp\left(-\frac{\beta\hbar\omega}{2}\right)-\exp\left(\frac{\beta\hbar\omega}{2}\right)}{\exp\left(-\frac{\beta\hbar\omega}{2}\right)+\exp\left(\frac{\beta\hbar\omega}{2}\right)}\tilde{C}''(\omega)\coth\left(\frac{\beta\hbar\omega}{2}\right)\sin\omega t\nonumber\\
 & =&-\frac{1}{\pi}\int_{0}^{\infty}d\omega\left(-\tanh\left(\frac{\beta\hbar\omega}{2}\right)\right)\tilde{C}''(\omega)\coth\left(\frac{\beta\hbar\omega}{2}\right)\sin\omega t\nonumber\\
 & =&\frac{1}{\pi}\int_{0}^{\infty}d\omega\tilde{C}''(\omega)\sin\omega t.
\end{eqnarray}


Thus, we can obtain the relationship between the real
and imaginary parts of the TCF as
\begin{equation}
C''(t)=\tan\left(\frac{\beta\hbar}{2}\frac{d}{dt}\right)C'\left(t\right).\label{eq:Trela}
\end{equation}
To conclude, the TCF can be rewritten as
\begin{equation}
C\left(t\right)=C'\left(t\right)+i\tan\left(\frac{\beta\hbar}{2}\frac{d}{dt}\right)C'\left(t\right).
\end{equation}

\end{widetext}

\providecommand{\noopsort}[1]{}\providecommand{\singleletter}[1]{#1}%

\end{document}